\documentclass{article}
\usepackage{graphicx} % Required for inserting images

\usepackage{url,hyperref,lineno,microtype,subcaption}
\usepackage[utf8]{inputenc}
\usepackage[T1]{fontenc}

\title{Making Online Polls More Accurate: Statistical Methods Explained}
\author{Alberto Arletti, Maria Letizia Tanturri \& Omar Paccagnella \\ 
Department of Statistical Science, University of Padua, Padua, Italy}
\usepackage{graphicx}   % For including graphics
\usepackage{natbib}     % for citation
\usepackage{listings}   % For code listings
\usepackage{amsmath}

\usepackage{tikz}       % for illustration
\usetikzlibrary{matrix, positioning} 

\begin{document}

\maketitle

\section*{Abstract}
Online data has the potential to transform how researchers and companies produce election forecasts. Social media surveys, online panels and even comments scraped from the internet can offer valuable insights into political preferences. However, such data is often affected by significant selection bias, as online respondents may not be representative of the overall population. At the same time, traditional data collection methods are becoming increasingly cost-prohibitive. In this scenario, scientists need instruments to be able to draw the most accurate estimate possible from samples drawn online. This paper provides an introduction to key statistical methods for mitigating bias and improving inference in such cases, with a focus on electoral polling. Specifically, it presents the main statistical techniques, categorized into weighting, modeling and other approaches. It also offers practical recommendations for drawing estimates with measures of uncertainty. Designed for both researchers and industry practitioners, this introduction takes a hands-on approach, with code available for implementing the main methods.

\section{Introduction}

One of the most powerful tools in the hands of researchers to investigate nature was introduced for the first time in 1934: random sampling. The idea is simple. Given a small portion of individuals in a group, it is possible to obtain a reliable estimate for the parameter of interest for the whole population, such as the population mean. A random sample, or probability sample - adjectives which will be used interchangeably in the text - possesses therefore the seemingly magical power of achieving an estimate even with few values of $n$, the sample size, compared to $N$, the population size \citep{smith1976foundations}. The key of such feat of random sampling lays in managing to obtain a sample which is \textit{entirely random} with respect to all aspects that might influence the parameter of interest. To do so, researchers often need to know the probability of each individual in the population to join the survey, a value called inclusion probability. If such value is known, and is not zero, the sample can be considered a  representative probability sample. \\

Although straightforward, respecting such requirement in practice can be a major issue. This is especially true when drawing measurements of complex human phenomena, such as voting behavior. As stated by \cite{kruskal1979representative} "the idea will rarely work in a complicated social problem because we always have additional variables that may have important consequences for the outcome" (p. 249). \\

The delicate complexity of social problems require random sampling to follow extra steps in order to obtain effective randomness in the sample, and therefore maintain its status as the "dominating" sampling mechanism. For example, when contacting citizens in order to measure their voting intentions for an upcoming election, randomness could be achieved by calling phone numbers at random, given a list of all phone addresses in a given area (method referred to as Random Digit Dialling, or RDD). But what about people who can't answer the phone, for a variety of reasons that \textit{could} be connected with their choice of vote, and therefore generate bias in the final outcome? In other words "polling of humans is far from the simple random sampling described in many statistics textbooks" \citep[p. 69]{gelman2021failure}.\\

The issue to achieve randomization when human factors are at play is furthered hampered by another important aspect: declining response rates. The decrease has been abundantly reported \citep{brick2011future}, with a recent example being the decline from 60\% in 2004 to 40\% in 2024 in the European Social Survey \citep{ESS2024}. This decline applies to electoral polls as well \citep{gelman2021failure}. As people are seemingly uninterested in answering field researchers, two consequences appear: Firstly, a non-response bias is introduced in electoral polls \citep{shirani2018disentangling}, which is to say that the individuals who do not respond could be systematically different from those who do respond. Secondly, conducing research becomes more expensive, as more and more people need to be contacted to obtain a representative sample \citep{baker2013summary}. Representative surveys can be also more expensive even without considering response rates. For example, \cite{unangst2020process} reports the cost for a single interview to be 10\$ when conveniently obtained from the internet, with little guarantees of randomness, while the cost climbs to 192\$ for the more selection-safe face-to-face approach. In addition to the prohibitive costs, given the complexity of social sciences and the increasing rates of non-response, one might legitimately question whether a truly random sample is still achievable at all. These two consideration led some researches to state that there is no such thing as a "random sample" anymore \citep{bailey2023new} or, humorously, that "non-random samples are almost everywhere" \citep[p. 718]{meng2018statistical}. These two consequences, have, in turn, caused researchers and polling companies to search for more convenient methods of sampling. Non-probability sampling is therefore introduced.

\subsection{Non-probability samples}

Given the aforementioned problems, researchers might be in need of alternative methods for data collection. To the rescue come non-probability samples, or non-random samples. Non-probability samples are all samples that come from a vast number of techniques used to obtain data, from snowball sampling \citep{dusek2015using} to asking people's opinion on social media \citep{alexander2020combining} to scraping web-pages \citep{schirripa2025inference}, to many others.  Such samples are cheaper and more convenient to obtain, and therefore very popular choice for researchers and practitioners. \\

In the social sciences, non-probability samples can be advantageous due to their versatility, low cost and possibility of being employed where other methods often cannot. In particular, speed can be a remarkable quality. For example, the influx of online non-probability data can allow feats such as using Facebook Advertising Platform to now-cast the distribution of migrants groups in the United States, as in \cite{alexander2020combining} and \cite{zagheni2017leveraging}. Non-probability samples can also be used to make updated forecasts when more recent census data is unavailable, such as in using Google searches to forecast birth rates \citep{billari2016forecasting}. Continuing, non-random sampling can often be the only viable strategy to examine hard-to-reach populations, as for example using mobile and landline in \cite{de2021design}, using LinkedIn as in \cite{dusek2015using} or in using the social media platforms Vkontakte e Odnoklassniki \citep{rocheva2022targeting}. Migrants are an especially salient case of such populations, which might not fit in the traditional administrative or random sampling schemes. For example, \cite{zagheni2014inferring} used localized tweets to draw a non-random sample used to infer migration patterns, while \cite{jacobsen2021using} used a tracking app for the same aim. Finally, it is clear that using social media, a case of non-probability sampling, offers the advantage of smaller prices and a relative large pool of individual to draw from.

\subsubsection{Pollings and the shortcoming of non-probability samples} \label{sec:goal}

Even though it is clear that non-probability samples such as online or social media data can be a game changer in many scenarios, the significant drawbacks of selection bias, which might results in less accurate results, must be accounted for \citep{callegaro2014critical}. Selection bias can be defined as systematic differences between the sampled and target populations, due to the fact that the survey was accessible to a section of the population only, for example internet users only or Facebook users only. Non-probability samples are non-representative as in they carry selection bias, which leads to a violation of the canon of randomness in some measure. 
Because non-probability samples contain this selection, drawn estimates, such as the predicted share of votes for a said party, are not reliable, as in they do not represent the target population of interest, but rather the selected subgroup from data was extracted (e.g. Facebook users who happened to be online at the time of the survey). Therefore, while selection is used in random sample to select a sample which is random in all its characteristics with respect to the interest statistics, non-random samples are vulnerable to the adverse effect of selection \cite[p. 246]{kruskal1979representative}. Some examples of such violations of the pure assumptions of probability sampling are non response, incomplete coverage of the population, and measurement errors \citep{brick2011future}. The effect is that the "magical" quality of random samples is not applicable anymore, and suddenly the small size $n$ of the sample is unable to measure correctly the large $N$ of the interest population \citep{meng2022comments}. Therefore, this has led the American Association for Public Opinion Research (AAPOR) in 2010 \citep{prepared2010research} and again in 2013 \citep[p. 12]{AAPOR2013} to state that "researchers should avoid non-probability opt-in-panels when a key research objective is to accurately estimate population values .. claims of representativeness should be avoided when using these sample sources".\\

Another problem is that respondents in non-probability surveys, such as ones collected via social media or online panels, exhibit less informative responses compared to more "involved" methods such as face-to-face. For example \citep{fricker2005experimental} and \citep{heen2014comparison} report "depressed responses" in the interviews, which shows for example in scores being centered around the middle of the scale and in less answer differentiation or smaller frequency of extreme opinions. \\

Arguably, the field where non-probability samples's shortfalls have generated the strongest shockwave is electoral polling \citep{evans2018value, zagheni2015demographic, shirani2018disentangling}. As put eloquently in a 2018 review: "Polls have had a number of high-profile misses in recent elections. Political polls have staggered from embarrassment to embarrassment in recent years" \citep[p. 757]{prosser2018twilight}. Famous examples are the 2016 presidential race \citep{kennedy2018evaluation} (which has been named "a black eye" for polling \citep[p. 67]{gelman2021failure}), the 2016 Brexit referendum \citep{FTBrexit2016} and the 2023 Turkish general elections \citep{selcuki2023turkishpolls}. Generally, the failure of those polls is mainly attributed to the use of non-probability samples \citep{gelman2021failure}, as such samples has been reported as less accurate compared to probability sources \citep{sohlberg2017determinants, sturgis2018assessment}. Nonetheless, the trend does not seem to be stopping for the rise of non-probability samples in electoral polling as well \citep{callegaro2014online}. It is clear that a failure in an electoral prediction bears a higher cost for the public image of the discipline. After all, "election polling is arguably the most visible manifestation of statistics in everyday life" \citep[p. 608]{shirani2018disentangling}. Election polling is almost most salient because poll-based forecasts are compared to actual election outcomes \citep{gelman2021failure}. \\

Researchers might end up stuck between a rock and a hard place. Random samples can hardly be completely trustworthy and require heavy costs compared to the cheaper non-probability alternatives \citep{tam2015big}. On the other hand, non-probability samples carry important challenges for inference. Given these premises, what should researcher do with the abundant quantities of non-random samples available, such as Twitter posts, Google searches, online and opt-in panels etc..? It is clear the need for reliable approaches to draw valuable inference from non-probability samples is pressing and might bring great benefits to the academic community. After all "Great advances of the most successful sciences - astronomy, physics, chemistry were and are, achieved without probability sampling. " \citep[pp. 28–29]{wiegand1968kish}.\\

From this scenario, the need for statistical methods used to draw valid inference from non-probability social science data emerges as paramount for the whole scientific community. Statistical methods could aim at reducing or acting as counterweight to the distortion or bias present in such non-probability samples. In other words, the estimated value would be closer to the true population value after applying the estimation method, in form of calibration or correction.\\

Given the potential of online and non-probability data for social sciences and opinion research, such as electoral polling, it is crucial to explore statistical methods that reduce bias and improve accuracy in such datasets. This work aims to assist researchers and practitioners by outlining key statistical techniques for correcting non-probability data, focusing on reducing distortion or bias. It provides an accessible overview of these methods, their assumptions, and practical implementation, serving as a reliable guide for selecting and applying the appropriate approach in their analyses.

\section{Data Availability Scenarios in Non-Probability Sampling} \label{sec:scope}  

Addressing selection bias in non-probability samples requires appropriate statistical methods, but their applicability depends on the available population information. Researchers may find themselves in different data availability scenarios when working with non-probability samples, which are here briefly illustrated. \\

In the simplest case, only sample data is available, with no population reference (e.g., hard-to-reach groups like migrants, where census data is lacking). More commonly, researchers also have population totals, as in electoral data, which may be available in marginal (e.g., total voters by sex or region) or cross-tabulated form (e.g., female voters by region). Lastly, some non-probability samples can be paired with a (often smaller-sized) probability sample \cite{tutz2023probability,rafei2022robust}. The present contribution focuses on the second case, where marginal or cross-tabulated totals are available. The first case allows little room for correction, while the third involves distinct challenges and is less common in electoral polls practice.\\

\begin{figure}
    \centering
    \includegraphics[width=0.5\textwidth]{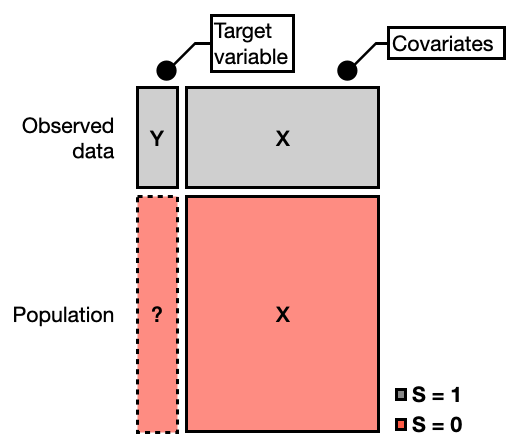}
    \caption{Schematic representation of the variables in the considered dataset.}
    \label{fig:dataset_schema}
\end{figure}

In the second setting, population information is available as either marginal totals or cross-tabulated census data. This can be represented as a dataset with a target variable $Y$, a set of covariates $X$ with $p$ parameters, and a $p$-sized vector $T(X)$ containing population totals for each variable in $X$. When complete cross-tabulated census data is available, the researcher has two datasets: 1. A non-representative sample containing $Y$ and covariates $X$ ($n$ rows).   2. A representative dataset of the full population ($N$ rows) with covariates $X$, but without $Y$. These datasets can be concatenated with an indicator variable $S$, where $S = 1$ for sampled units and $S = 0$ otherwise (see Figure \ref{fig:dataset_schema}).  \\

An additional important concepts is population cells. Any population, such as voters in a country, can be divided into non-overlapping cells. Each cell represents a unique category in the population, defined by a specific combination of categorical $X$ variables. For example, a cell might be male, 30-45 years old, voter. The total number of cells is given by the product of the levels of available categorical variables. For instance, if gender (2 levels) and employment status (3 levels) are available, the population is divided into $2 \times 3 = 6$ cells.  \\

Finally, hands-on practice enhances the learning of new methods. To complement the theoretical discussion, this introduction is accompanied by a sample dataset and code implementations for most methods presented. This allows readers to grasp both the technical details and practical application. The code and data are available on GitHub: \href{https://github.com/alberto-arletti/nonign_sel_companion}{\texttt{nonign\_sel\_companion}}.

\section{Weighting} \label{sec:weighting}

Weighting is considered one of the most important methods for correcting a non-representative sample \citep{valliant2020comparing}. In weighting, the individual observations are up or down weighted so their distribution is adapted to be more similar to the distribution of a representative sample or of the census. In their most basic idea, if the sample has way more males than females compared to the known national totals, then, male observations can be down-weighted. This class of methods can also be referred to as "pseudo-weighting" or "quasi-randomization" \citep{Vaillant_2019}. This is due to the fact that in random sampling, observations in the sample are weighted by the inverse of their inclusion probability, which is known (see \cite{horvitz1952generalization}). In the case of non-random sampling, the inclusion probabilities are not known, and are to be estimated. Therefore, weighting is used in trying to approximate sampling weights in a manner which is resembling that what is done in probability sampling. In the case of unknown inclusion probabilities, or non-random samples, weighing can be obtain with one, or a combination of raking, propensity scoring and matching. 

\subsection{Raking (Iterative Proportional Fitting)} \label{Raking}
Iterative Proportional Fitting, or Raking, \citep{raking40} is a weighting method which is used to weight a dataframe so that the $X$ variables marginals match the corresponding population marginals. This is done in the case of multiple marginal distributions, for example gender and region. The term iterative is used to refer to the process which is used to obtain the weighting, which can be described in simple words as adjusting the weights iteratively, making them more similar to the marginals at each iteration until convergence \citep{stephan1942iterative}. 

The goal of raking is to assign weights $w_1 \dots w_j \dots w_n$ to each row in the sample so that the weighted sums match known population totals from the census. For the covariate $p$, This can be expressed as:  
\begin{align} % \tag{raking}
    \sum_{j = 1}^n w_j x_{j,p} = T(X_p). 
\end{align}  
Here, $T(X_p)$ is the population total for the $p$-th covariate, and $x_{j,p}$ represents the value of the $p$-th covariate for row $j$. The estimated population mean ($\widehat{\mu}(Y)$) is then obtained as:  
\begin{align} \label{eq:raking_mean}
    \widehat{\mu}(Y) = \frac{1}{n} \sum_{j = 1}^n y_j w_j.
\end{align}

This formula allows the estimation of the population total for the target variable, such as share of votes. If one would like to obtain measures of uncertainty around such estimate, a common practice is to use a bootstrap or similar resampling approaches \citep{kolenikov2010resampling}. \\

Raking is a very simple weighting method that only requires the marginal distributions to me employed, and is especially useful in the case where marginals totals only are available (see Section \ref{sec:scope}), or in cases where the number of observations in each cell is small. Nonetheless, it can suffer from a series of limitations. To begin with, the raking to the marginals does not take into account possible higher level interactions between the raking variables. This can be an issue making the weighing less accurate compared to the real population distribution. A proposed solution to this problem is multilevel calibration weighting, an approach by \cite{ben2021multilevel}. While raking only match the marginal distributions of the raked variables, it might not not be able to balance higher-order interactions. Multilevel calibration weighting aims at solving that, behaving similarly to raking but adding with some approximate balance for interaction, prioritizing lower order interactions. In addition, if the raking variables do not fully account for the inclusion probability, the method becomes inconsistent. Finally, it should be noted that the weights produced by raking can have very high (or very low) values, making the practice unreliable.  One possible solution is trimming the weights, a solution which is also implemented directly in the R command for raking \texttt{anesrake} \citep{pasek2018package}.

\subsection{Propensity Score Adjustment} \label{IPW}

Propensity Score Adjustment is a class of adjustment methods which rely on the estimation of the probability of inclusion in the non-probability sample. The main method, discussed here, is often referred as Propensity Score-based Inverse Probability Weighting (PS-IPW) \citep{zou2016variance}.\\

PS-IPW works through the use of a second representative sample, with common covariates to the non-probability sample, there the target $Y$ variable is missing \citep{10.1214/16-STS597, mcpheedata}. Such sample can be generated knowing the census cross-tabulated totals, if those are available. To do so, it would be sufficient to generate a dataframe where each column corresponds to a census cross-tabulated variable, and the number of rows that belong in each cell correspond to, or are proportional to, the know population total. The two datasets are temporarily binned into a single frame, as described in Section \ref{sec:scope} and Figure \ref{fig:dataset_schema}. Then, the method builds a weighted logistic regression model to estimate the probability of an observation being in a non-probability sample. Here, the regression weights correspond to the known inclusion probabilities in the reference sample, while non-sampled observations receive a weight of 1. Inclusion probabilities in the reference sample correspond to the known probability of individual $j$ of being included in the sample, which generally accompany a representative sample. If the reference sample has been generated from the cross-tabulated census values, then the inclusion probability of a row $j$ belonging to cell $c$ is the just the inverse of the numerosity of that cell. The regression can be described as 
\begin{align*}
   \text{logit}(P(S = 1 | X)) = \beta_0 + \beta^T X.
\end{align*} 
The predicted values of the weighted regression, which can be set as $\hat{\pi}_j = \text{logit}(P(S_j = 1 | x_j))$, are then inverted and used to estimate the population mean.  
\begin{align} \label{eq:PS-IPW} % \tag{PS-IPW} 
    \widehat{\mu}_{\text{PS-IPW}}(Y) = \frac{1}{n} \sum_{j = 1}^n \frac{1}{\hat{\pi}_j} y_j. 
\end{align} 

This last formula is the same of the famous Horvitz-Thompson estimator \citep{horvitz1952generalization}, with the difference that the weights are not known from the sample design, but are estimated from the data. It is also similar to Equation \ref{eq:raking_mean}, with the difference that $\frac{1}{\hat{\pi}_j}$ are estimated differently from $w_j$. Such probability estimated from the data is called the propensity score. What this achieves is an estimation of the inclusion probabilities, which are unknown, from the observed data. The propensity score represents the conditional probability of being included in the survey given an individual's covariate profile.\\

For a measure of uncertainty of this estimate, variance estimates can be obtained through a Taylor linearization approximation \citep[p. 426]{valliant2013practical} or through a Jackknife approximation \citep[p. 8]{Vaillant_2019}. \\

One topic of discussion regard which model to choose in order to obtain propensity scores. While logistic regression is a very popular method, some authors argue that it is insufficient in cases such as where the propensity score shows a non-linear function. On this regard, \cite{lee2010improving} compare the performance of different methods to obtain propensity scores. They compare both logistic regression with Classification and Regression Trees (CART) models, and find that logistic regression's performance can deteriorate in case of non-additivity and non-linearity. Therefore, choosing a flexible or non-parametric approach to model propensity scores can be advantageous.\\

Once an appropriate model has been selected to capture the selection mechanism, and that there are no empty cells in the data, the PS-IPW can be used to build reliable estimators. A first assumption of this approach is that every unit in the population has a non-zero propensity score. A second important assumption is that the covariates $X$ should include all relevant confounders \citep{lee2006propensity}. The main danger of this method is therefore in the case there the selected $X$ variables do not fully account for the sample selection mechanism, or in other words, the there is significant selection bias which cannot be controlled by the available covariates. In that case, adjusting for the propensity score will not produce unbiased estimates of the treatment effect. A further requirement of PS-IPW is called "common support" and requires that the distribution of the covariates in the reference sample is similar to the distribution in the sample to adjust. For example, there should not be population cells completely absent from the non-probability sample \citep{Vaillant_2019}. Pseudo-inclusion probabilities are typically estimated using weighted logistic regression \citep{lee2006propensity}. 

\section{Modelling} \label{sec:modelling}

Another popular approach to adjust non-probability surveys and remove selection bias is modelling. In this case, the non-random sample is employed to train a model used to predict the dependent variable for each cell of the missing rows, corresponding to the population. This approach is also called superpopulation model estimation \citep{Vaillant_2019} or model-based estimation \citep{wu2022statistical}. In modelling, the $y_i$ values of the non-sampled units are predicted with a variety of methods trained on the sampled units. In this way, the value for the total population is considered the union of both the sampled and non-sampled units. That is, the non-sampled units correspond to all individuals who are in the target population, but not in the sample. 

\subsection{Post-stratification}

Superpopulation methods are therefore comprised of two steps, a first modelling step, where the model is estimated from the observed data, and a post-stratification step, where the value is predicted for each cell of the population. The sum of all predicted values for all cell gives the estimated value for the entire population. After modelling, the post-stratification step allows to balance for sample discrepancy. \\

An estimate of the population mean for a given cell of the population, $y_c$, where the subscript $c$ indicates the cell, can be therefore obtained by first estimating a model between $Y$ and $X$ in the sample, for example, a linear regression. 
This can be described as: 
\begin{align} \label{eg:modelling} % \tag{modelling} 
    \text{logit}(P(Y = 1 | X)) = \beta_0 + \beta^T X.
 \end{align}  
Then, the cell total can be obtained using the following formula: 
\begin{align} \label{eq:post-stratification} % \tag{post-stratification} 
    \widehat{\mu}(Y_c) = \frac{N_c}{N} (\beta_0 + \beta^T X_c)  
\end{align} 
where $N_c$ is the known size of the population cell $c$, $X_c$ indicates the matrix of covariates data for the cell $c$, and $\widehat{\beta}$ are the estimated regression coefficients. The estimate for the whole population will be the sum of all cell totals, so that $\widehat{\mu}(Y) = \sum_c \widehat{\mu}(Y_c)$. \\

While the post-stratification adjustment step remains the same across application, what can be changed is the model used for prediction. A simple linear model can be substituted with more complicated or non-linear models. In a very famous application, a hierarchical model is used, giving rise to Multilevel Regression and Post-stratification (MRP). MRP is not of recent development, with \cite{gelman1997poststratification} being the original proposer of the method. Nonetheless, MPR is one superpopulation method that is frequently used with online surveys \citep{mcpheedata, si2020use}. In MRP a multilevel regression model is used to estimate the outcome variable using a larger number of auxiliary variables and their interactions than is possible with standard weighting methods. The particularity of MRP is that it performs a cell-based (sub-group) estimation, and the hierarchical component (with Bayesian prior in its original specification, see \cite{li2022embedded}) regularizes the model and allows for borrowing of information. \\

MRP is a key method in the field, and it provides several advantages over a post-stratification with a simple linear regression. To best understand the mechanics of MPR, it can be useful to examine the following formula for estimating the population mean using MRP \citep[p. 5]{si2020use}:
\begin{align} \label{eq:MRP} % \tag{MRP} 
    \widehat{\mu}_{\text{MRP}}(Y) = \sum_{c} \frac{N_c}{N} \frac{\widehat{\mu}(Y_c) + \delta_c \widehat{\mu}(Y)}{1 + \delta_c}, \quad \text{where} \quad \delta_c = \frac{\sigma^2_c}{n_c \sigma^2_Y}.
\end{align}
Here, as in Formula \ref{eq:post-stratification} the subscript $c$ indicates a  post-stratification cell, $\widehat{\mu}(Y_c)$ is the model estimate for cell $c$, $N_c$ is the size of cell $c$ in the population, $\widehat{\mu}(Y)$ is the estimated population mean, $\sigma_c^2$ is the variance of the outcome variable for cell $c$, $n_c$ is the sample size for cell $c$ and $\sigma^2_Y$ is the outcome between-cell variance. Between-cell variance is a measure of how much the mean of $Y$ differs from one cell to another, reflecting systematic differences between groups defined by the stratifying variables (e.g., age, gender, region). Therefore, what this formula tells us, is that the less information we have on cell $c$, both in terms of sample size and variety, the more we are going to "borrow" from the other cells. This mechanics is especially effective in online samples or social media samples, where it's quite often the case to have cells with very few observations. \\

For uncertainty measures on the population estimates of both post-stratification and MRP usually a Bayesian approach with posterior draws is preferred \citep{lopez2022multilevel}.\\

It has been noted that post-stratification is useful in reducing selection bias and for correcting imbalances in the sample composition. One advantage of such estimators is their ability to reduce bias \citep{kim2021combining}. On this regard, the method has shown to be capable of impressive bias-correcting performances in election forecasting, for example in \cite{wang2015forecasting}. Nonetheless, when drawing inference with such method some factors come into play to determine its performance. The first factor is the need for high quality predictive post-stratification variables, or, in other words, variables with strong relationship with the outcome variable. Authors have reported how poorly predictive auxiliary information might have an important effect on the final outcome \citep{si2020use}, and that variables chosen for post-stratification being more relevant than the model used for estimation \citep{prosser2018twilight}. For example, \cite{buttice2013does} examines the correlates of MRP performance in various scenarios. The authors examine how MRP accuracy of estimates of election results varies as as the strength of relationship between voting opinion and state-level covariates increases. They observe that as strength of the relationship between opinion and the state-level covariates increases, then also MRP estimates gets closer to the true values. This is not seen with the same strength for the individual level covariates.\\

The requirement for high quality post-stratification variables can be challenging when the census is limited. The requirement to have cross-tabulated population tables can be daunting especially as the number of covariates increase. Therefore, it is often the case that variables useful for adjustment are not included in the census, such as party identification or previous vote \citep{gelman2021failure}. Usually, due to non-availability in census, post-survey adjustments are limited to basic demographics such as age, gender, race and education from large-scale government surveys \citep{chen2019calibrating}. Moreover, for the case of electoral polling, these problems can be exacerbated for practitioners working outside of the United States. For example, pollsters in the United States can access party registration information, which is generally unavailable in other countries \citep{prosser2018twilight}. Therefore, MRP has, generally been applied so far in election forecast for a few countries \citep{leemann2017extending}. As a possible solution, \cite{kastellec2015polarizing} suggest expanding the post-stratification table by incorporating a survey that includes one or more non-census variables, which can aid in adjusting for discrepancies between the sample and the target population. Such practice, can be referred to as "embedded MRP" or e-MRP \citep{li2024embedded}.

\section{Other Methods} \label{Other Methods}

\subsection{Sample Matching}
Matching is a technique that can be both applied before the sample is selected \citep{cornesse2020review, bethlehem2016solving} or after the non-probability sample is already obtained \citep{mercer2018weighting}. The approach for the second case, the one of interest for the purpose of the present work, is attributed to \cite{rivers2007sampling}.  Similarly to Propensity Score Adjustment, it requires a probability sample where the target variable does not need to be measured, but where there are matching covariates. The reference sample is treated as a target, where each row of the target is paired with the closest observation in the non-probability sample. The "matching" observation is chosen to be an observation which has the strongest similarity in the covariates. An euclidean distance metric can be used \citep{cornesse2020review}, as well as any sort of similarity matrix such as the one in a Random Forest \citep{mercer2018weighting}, or a nearest neighbor approach which can especially useful in the case of continuous variables, or categorical variables with many ordinal levels, where all combinations of the covariates might not be practical \citep{chen2000nearest}. The closest match is chosen for each row of the reference sample, and any remaining observation which has not been paired is discarded. Sequentially, each observation in the target dataframe is matched one at the time, and the most similar case is chosen among the cases which has not been matched previously. Then, the statistics of interests are obtained using the target variable $y$ of the matched cases. In other words, each row of the target reference sample is substituted with the most similar observation in the non-probability case. \\

The main limitation of matching is that, in order to obtain a meaningful matching, a sufficiently large set of variable should be available in the required probability sampling. Most often, these variables should be different than the common demographic variables and might not be present in available census. Otherwise, other forms of adjustment would be more straightforward. For the case of electoral polling, obtaining a reference sample with such characteristics can be challenging. 

\subsection{Inverse Sampling}
Inverse Sampling is presented for the estimation of non-probability big data samples in \cite{kim2019sampling}. The idea of inverse sampling is to leverage on the large $n$ of the non-probability sample to make a sub-selection. The first-phase sample consists of big data, named $A$, which is affected by selection bias. The second-phase sample, named $A_2$ is a subset of the first-phase sample, designed to adjust for this selection bias. To extract the subsample, inclusion probabilities proportional to the importance weights are used for selection. External information from a reference sample or from census are used to correct for selection bias in the second step. 

\subsection{Doubly-Robust Post-stratification} 

Doubly-Robust Post-stratification (DRP) is substantially a combination between weighting, seen in Section \ref{sec:weighting}, and modelling, seen in Section \ref{sec:modelling}. The fundamental idea is to combine the two components, a propensity score component and a modelling with post-stratification component. When estimating a propensity score model, the specified model might be incorrect, for example, it might ignore interactions which are influencing the selection mechanism. The same might be for the modelling approach, where the chosen model might not be the best fit to describe the relationship between the target variable ($Y$) and the available covariates ($X$) \citep{tan2007comment}. In DRP, the final estimate will be correct as the sample size increases even if one of the two models, either the modelling or the propensity score model, is incorrect or misspecified (Theorem 2, \cite{chen2020doubly}). This guarantees further protection against bias. Similarly to DR-IPW, imagine a second reference sample where $Y$ is missing is available. We call the non-probability sample $A$ and the reference probability sample $B$. To obtain DRP, two models are fitted, \begin{enumerate}
    \item A propensity score model on the probability of $j$-th being included in $A$, using, for example, a weighted logistic regression as in Equation \ref{eq:PS-IPW}. The predicted propensity score is again $\hat{\pi}_j$ for row $j$. 
    \item A model of the relationship between the target $Y$ and the covariates $X$, using $A$ data only, as in Equation \ref{eg:modelling}. The predicted value of $y$ for row $j$ by this model is indicated as $\hat{y}_j$. 
\end{enumerate} 
The final DRP population estimate is obtained by: 
\begin{align} \label{eq:DRP} % \tag{DRP} 
    \hat{\mu}_{\text{DR}}(Y) = \frac{1}{\sum_{i \in A} 1 / \hat{\pi}_i} \sum_{i \in A} \frac{1}{\hat{\pi}_i} ( y_i -  \hat{y}_j) + \frac{1}{N} \sum_{c} N_c (\beta_0 + \beta^T X_c). % \hat{y}_j.
\end{align}
Unpacking this expression, $\sum_{i \in A}$ indicates to sum across all rows in the $A$ dataframe, while $\sum_{c}$ for each population cell. The first term in \ref{eq:DRP} sums the difference between the predicted and the measured values of $Y$ for the non-probability sample, weighted by the inverse of the probability score obtained with the propensity score model. The second term is a post-stratification, as in Equation \ref{eq:post-stratification}. \\

An estimate of variance of the DRP is present in \cite{chen2020doubly}, but bootstrap resampling can also be used (eg. see \cite{berkesewicz2024inference}). Point estimation of DRP in R can be easily carried out with the \texttt{nonprobsvy} package \citep{nonprobsvy}, which also provide estimates of the uncertainty of the predicted population mean. The method has a set of strong qualities on paper but real-world application might vary widely (Si, J. personal communication, 11/2023). This might be due to the fact that the two model components' effect might interact with one another creating either a more unpredictable behavior (Meng, X. L., personal communication, 11/2023). In conclusion, DRP offers notable advantages theoretically but not yet replace other methods in practical applications automatically. 

\section{Limits of the presented approaches}

All the models presented in the previous sections assume that the selection mechanism to be entirely explained by the $X$ covariates alone. If the selection mechanism is not entirely explained by $X$, then the estimated model might not provide accurate estimates of the population of interest. Importantly, there is abundant evidence that non-probability samples might suffer from non-ignorable selection, or in other words, that $S$ is not only influenced by $X$ alone but by the target variable $Y$ as well. In political polling, this might be the case due to a variety of reasons. For example, respondents in non-probability panels being generally more politically engaged than the general population \citep{prosser2018twilight}, respondents who vote for a candidate who is doing well might be more likely to answer a survey \citep{gelman2016mythical}, ads being used to target responders failing to be neutral or to attract voters of a specific political affiliation \citep{matz2017psychological, zarouali2022using, schneider2022s, kuhne2020using}, online responders having different personality characteristics compared to the global population \citep{valentino2020consequences, bruggen2010determinants}, or online samples having no respondents in certain cells of the population \citep{mixmode19demetra}. Notwithstanding the multiplicity of mechanisms which might induce non-ignorability in an online sample, there is lack of methods able to take into account such issue. One field where methods might be applied to this case is missing data theory, where mechanism missingness can be considered the same driving selection, simply inverted. In this case, some reweighing methods have been proposed to adjust for non-ignorable missingness (for examples, see \cite{matei2018some}), as well as models which use assumptions on the selection mechanism to adjust for selection bias (see \cite{west2023evaluating} and \cite{andridge2024using}). \\ 

All in all, while the methods here presented might provide capable in removing selection bias from samples collected online, researchers should be conscious that some selection mechanisms cannot be completely undone without stronger assumptions or knowledge on the sampling mechanism. 

\section{conclusions}

This paper reviewed the main methods used for adjusting a non-probability sample such as an online sample, with focus on electoral polling. While each method has been described in general terms, the choice of which one to use in each situation can depend on the specific setting, data availability and research goal. One useful resource in this regard is \cite{cornesse2020review}, which also had a setting centered on non-probability samples used to estimate election polls. The authors compare probability samples with corrected or weighted non-probability samples. They compare some approaches listed in the previous sections: a) Calibration weighting using post-stratification or raking; b) Sample matching; c) Propensity score weighting; d) Pseudo-design based estimation such as propensity score weighting; They find that weighting can reduce the bias in some cases, but in general the authors arrived to the conclusion that weighting does not suffice in completely eliminating bias in non-probability based surveys. \\

One general rule that can be applied to all methods, is that as long as strong predictive variables are available, in weighting or in modelling alike, the most of the selection mechanism can be accounted for. As $X$ decreases in predictive power, things get more complicated: selection might be unaccounted for and researchers have less tools at their disposal in obtaining an estimate. Concluding, we go back to an important concept expressed in the introduction: the large $n$ (typical of non-probability samples), is, alone, unable to provide unbiased estimation. Nonetheless, a rich $X$, or a wide dataset of covariates might instead be a more fruitful pathway towards robust estimation. In this sense, to work well non-probability online samples should not just be big, but rich as well. Techniques that might be the most promising in this sense are therefore the ones which allow for an expansion of prediction variables such as in \cite{li2024embedded, kuriwaki2024geography}, and methods that allow the researcher to add previous knowledge on the possible selection mechanism, such as in \cite{little2020measures}. In general, estimation with non-probability samples in electoral polling should proceed carefully depending on the selection mechanism. \\

While non-probability samples pose significant challenges due to selection bias, they also offer valuable opportunities when handled with the right statistical methods. This paper has provided both an intuitive and technical overview of key approaches to adjust for bias and improve inference. Although no method can fully replace probability sampling, the techniques discussed here can enhance the reliability of estimates derived from non-representative data. By increasing awareness of both the risks and potential of these samples, this work aims to support researchers in making informed methodological choices when working with online and other non-probability datasets. 

\section*{Funding}
This work was part-funded by PON "Research and Innovation" 2014-2020 Actions IV.4 "PhDs and research contracts on innovation issues" and Action IV.5 "PhDs on Green issues." - Ministerial Decree 1061/2021 as a PhD studentship to Alberto Arletti. The funders had no role in study design, data collection and analysis, decision to publish, or preparation of the manuscript.

\section*{Data Availability Statement}

The code and data to reproduce the methods here indicated are available on the GitHub repository \href{https://github.com/alberto-arletti/nonign_sel_companion}{\texttt{nonign\_sel\_companion}}. 

\bibliographystyle{apalike}
\bibliography{main}

\begin{thebibliography}{}

\bibitem[Alexander et~al., 2020]{alexander2020combining}
Alexander, M., Polimis, K., and Zagheni, E. (2020).
\newblock Combining social media and survey data to nowcast migrant stocks in the united states.
\newblock {\em Population Research and Policy Review}, pages 1--28.

\bibitem[Andridge, 2024]{andridge2024using}
Andridge, R.~R. (2024).
\newblock Using proxy pattern-mixture models to explain bias in estimates of covid-19 vaccine uptake from two large surveys.
\newblock {\em Journal of the Royal Statistical Society Series A: Statistics in Society}, page qnae005.

\bibitem[Bailey, 2023]{bailey2023new}
Bailey, M.~A. (2023).
\newblock A new paradigm for polling.
\newblock {\em Harvard Data Science Review}, 5(3).

\bibitem[Baker et~al., 2010]{prepared2010research}
Baker, R., Blumberg, S.~J., Brick, J.~M., Couper, M.~P., Courtright, M., Dennis, J.~M., Dillman, D., Frankel, M.~R., Garland, P., et~al. (2010).
\newblock Research synthesis: Aapor report on online panels.
\newblock {\em Public Opinion Quarterly}, 74(4):711--781.

\bibitem[Baker et~al., 2013a]{AAPOR2013}
Baker, R., Brick, J.~M., Bates, N.~A., Battaglia, M., Couper, M.~P., Dever, J.~A., Gile, K.~J., and Tourangeau, R. (2013a).
\newblock Report of the {AAPOR} task force on non-probability sampling.
\newblock Technical report, American Association for Public Opinion Research (AAPOR).

\bibitem[Baker et~al., 2013b]{baker2013summary}
Baker, R., Brick, J.~M., Bates, N.~A., Battaglia, M., Couper, M.~P., Dever, J.~A., Gile, K.~J., and Tourangeau, R. (2013b).
\newblock Summary report of the aapor task force on non-probability sampling.
\newblock {\em Journal of survey statistics and methodology}, 1(2):90--143.

\bibitem[Bartoli et~al., 2019]{mixmode19demetra}
Bartoli, B., Fornea, M., and Respi, C. (2019).
\newblock Selection bias and representation of research samples: The effectiveness of mixing mode and sampling frames.
\newblock Poster presented at the 2019 GOR conference.

\bibitem[Ben-Michael et~al., 2021]{ben2021multilevel}
Ben-Michael, E., Feller, A., and Hartman, E. (2021).
\newblock Multilevel calibration weighting for survey data.
\newblock {\em Political Analysis}, pages 1--19.

\bibitem[Beręsewicz and Szymkowiak, 2024]{berkesewicz2024inference}
Beręsewicz, M. and Szymkowiak, M. (2024).
\newblock Inference for non-probability samples using the calibration approach for quantiles.
\newblock {\em arXiv preprint arXiv:2403.09726}.

\bibitem[Bethlehem, 2016]{bethlehem2016solving}
Bethlehem, J. (2016).
\newblock Solving the nonresponse problem with sample matching?
\newblock {\em Social Science Computer Review}, 34(1):59--77.

\bibitem[Billari et~al., 2016]{billari2016forecasting}
Billari, F., D'Amuri, F., and Marcucci, J. (2016).
\newblock Forecasting births using google.
\newblock In {\em Carma 2016: 1st international conference on advanced research methods in analytics}, pages 119--119. Editorial Universitat Polit{\`e}cnica de Val{\`e}ncia.

\bibitem[Brick, 2011]{brick2011future}
Brick, J.~M. (2011).
\newblock The future of survey sampling.
\newblock {\em Public Opinion Quarterly}, 75(5):872--888.

\bibitem[Br{\"u}ggen and Dholakia, 2010]{bruggen2010determinants}
Br{\"u}ggen, E. and Dholakia, U.~M. (2010).
\newblock Determinants of participation and response effort in web panel surveys.
\newblock {\em Journal of Interactive Marketing}, 24(3):239--250.

\bibitem[Buttice and Highton, 2013]{buttice2013does}
Buttice, M.~K. and Highton, B. (2013).
\newblock How does multilevel regression and poststratification perform with conventional national surveys?
\newblock {\em Political analysis}, 21(4).

\bibitem[Callegaro et~al., 2014a]{callegaro2014online}
Callegaro, M., Baker, R.~P., Bethlehem, J., G{\"o}ritz, A.~S., Krosnick, J.~A., and Lavrakas, P.~J. (2014a).
\newblock {\em Online panel research: A data quality perspective}.
\newblock John Wiley \& Sons.

\bibitem[Callegaro et~al., 2014b]{callegaro2014critical}
Callegaro, M., Villar, A., Yeager, D., and Krosnick, J.~A. (2014b).
\newblock A critical review of studies investigating the quality of data obtained with online panels based on probability and nonprobability samples1.
\newblock {\em Online Panel Research: Data Quality Perspective, A}, pages 23--53.

\bibitem[Chen and Shao, 2000]{chen2000nearest}
Chen, J. and Shao, J. (2000).
\newblock Nearest neighbor imputation for survey data.
\newblock {\em Journal of official statistics}, 16(2):113.

\bibitem[Chen et~al., 2019]{chen2019calibrating}
Chen, J. K.~T., Valliant, R.~L., and Elliott, M.~R. (2019).
\newblock Calibrating non-probability surveys to estimated control totals using lasso, with an application to political polling.
\newblock {\em Journal of the Royal Statistical Society Series C: Applied Statistics}, 68(3):657--681.

\bibitem[Chen et~al., 2020]{chen2020doubly}
Chen, Y., Li, P., and Wu, C. (2020).
\newblock Doubly robust inference with nonprobability survey samples.
\newblock {\em Journal of the American Statistical Association}, 115(532):2011--2021.

\bibitem[Cornesse et~al., 2020]{cornesse2020review}
Cornesse, C., Blom, A.~G., Dutwin, D., Krosnick, J.~A., De~Leeuw, E.~D., Legleye, S., Pasek, J., Pennay, D., Phillips, B., Sakshaug, J.~W., et~al. (2020).
\newblock A review of conceptual approaches and empirical evidence on probability and nonprobability sample survey research.
\newblock {\em Journal of Survey Statistics and Methodology}, 8(1):4--36.

\bibitem[De~Vries et~al., 2021]{de2021design}
De~Vries, L., Fischer, M., Kroh, M., K{\"u}hne, S., and Richter, D. (2021).
\newblock Design, nonresponse, and weighting in the 2019 sample q (queer) of the socio-economic panel.
\newblock {\em SOEP Survey Papers}, 940.

\bibitem[Deming and Stephan, 1940]{raking40}
Deming, W.~E. and Stephan, F.~F. (1940).
\newblock On a least squares adjustment of a sampled frequency table when the expected marginal totals are known.
\newblock {\em The Annals of Mathematical Statistics}, 11(4):427--444.

\bibitem[Dusek et~al., 2015]{dusek2015using}
Dusek, G., Yurova, Y., and Ruppel, C.~P. (2015).
\newblock Using social media and targeted snowball sampling to survey a hard-to-reach population: A case study.
\newblock {\em International Journal of doctoral studies}, 10:279.

\bibitem[{European Social Survey}, 2024]{ESS2024}
{European Social Survey} (2024).
\newblock Modes of data collection: The ess move to self-completion data collection.
\newblock \url{https://europeansocialsurvey.org/methodology/methodological-research/modes-data-collection}.
\newblock Retrieved October 16, 2024.

\bibitem[Evans and Mathur, 2018]{evans2018value}
Evans, J.~R. and Mathur, A. (2018).
\newblock The value of online surveys: A look back and a look ahead.
\newblock {\em Internet research}, 28(4):854--887.

\bibitem[{Financial Times}, 2016]{FTBrexit2016}
{Financial Times} (2016).
\newblock Brexit poll tracker.
\newblock Retrieved October 11, 2024.

\bibitem[Fricker et~al., 2005]{fricker2005experimental}
Fricker, S., Galesic, M., Tourangeau, R., and Yan, T. (2005).
\newblock An experimental comparison of web and telephone surveys.
\newblock {\em Public Opinion Quarterly}, 69(3):370--392.

\bibitem[Gelman, 1997]{gelman1997poststratification}
Gelman, A. (1997).
\newblock Poststratification into many categories using hierarchical logistic regression.
\newblock {\em Survey methodology}, 23:127.

\bibitem[Gelman, 2021]{gelman2021failure}
Gelman, A. (2021).
\newblock Failure and success in political polling and election forecasting.
\newblock {\em Statistics and Public Policy}, 8(1):67--72.

\bibitem[Gelman et~al., 2016]{gelman2016mythical}
Gelman, A., Goel, S., Rivers, D., Rothschild, D., et~al. (2016).
\newblock The mythical swing voter.
\newblock {\em Quarterly Journal of Political Science}, 11(1):103--130.

\bibitem[Heen et~al., 2014]{heen2014comparison}
Heen, M., Lieberman, J.~D., and Meithe, T. (2014).
\newblock A comparison of different online sampling approaches for generating national samples.

\bibitem[Horvitz and Thompson, 1952]{horvitz1952generalization}
Horvitz, D.~G. and Thompson, D.~J. (1952).
\newblock A generalization of sampling without replacement from a finite universe.
\newblock {\em Journal of the American statistical Association}, 47(260):663--685.

\bibitem[Jacobsen and K{\"u}hne, 2021]{jacobsen2021using}
Jacobsen, J. and K{\"u}hne, S. (2021).
\newblock Using a mobile app when surveying highly mobile populations: Panel attrition, consent, and interviewer effects in a survey of refugees.
\newblock {\em Social Science Computer Review}, 39(4):721--743.

\bibitem[Kastellec et~al., 2015]{kastellec2015polarizing}
Kastellec, J.~P., Lax, J.~R., Malecki, M., and Phillips, J.~H. (2015).
\newblock Polarizing the electoral connection: partisan representation in supreme court confirmation politics.
\newblock {\em The journal of politics}, 77(3):787--804.

\bibitem[Kennedy et~al., 2018]{kennedy2018evaluation}
Kennedy, C., Blumenthal, M., Clement, S., Clinton, J.~D., Durand, C., Franklin, C., McGeeney, K., Miringoff, L., Olson, K., Rivers, D., et~al. (2018).
\newblock An evaluation of the 2016 election polls in the united states.
\newblock {\em Public Opinion Quarterly}, 82(1):1--33.

\bibitem[Kim et~al., 2021]{kim2021combining}
Kim, J.~K., Park, S., Chen, Y., and Wu, C. (2021).
\newblock Combining non-probability and probability survey samples through mass imputation.
\newblock {\em Journal of the Royal Statistical Society Series A: Statistics in Society}, 184(3):941--963.

\bibitem[Kim and Wang, 2019]{kim2019sampling}
Kim, J.~K. and Wang, Z. (2019).
\newblock Sampling techniques for big data analysis.
\newblock {\em International Statistical Review}, 87:S177--S191.

\bibitem[Kolenikov, 2010]{kolenikov2010resampling}
Kolenikov, S. (2010).
\newblock Resampling variance estimation for complex survey data.
\newblock {\em The Stata Journal}, 10(2):165--199.

\bibitem[Kruskal and Mosteller, 1979]{kruskal1979representative}
Kruskal, W. and Mosteller, F. (1979).
\newblock Representative sampling, iii: The current statistical literature.
\newblock {\em International Statistical Review/Revue Internationale de Statistique}, pages 245--265.

\bibitem[K{\"u}hne and Zindel, 2020]{kuhne2020using}
K{\"u}hne, S. and Zindel, Z. (2020).
\newblock Using facebook and instagram to recruit web survey participants: A step-by-step guide and application.
\newblock {\em Survey Methods: Insights from the Field (SMIF)}.

\bibitem[Kuriwaki et~al., 2024]{kuriwaki2024geography}
Kuriwaki, S., Ansolabehere, S., Dagonel, A., and Yamauchi, S. (2024).
\newblock The geography of racially polarized voting: calibrating surveys at the district level.
\newblock {\em American Political Science Review}, 118(2):922--939.

\bibitem[Lee et~al., 2010]{lee2010improving}
Lee, B.~K., Lessler, J., and Stuart, E.~A. (2010).
\newblock Improving propensity score weighting using machine learning.
\newblock {\em Statistics in medicine}, 29(3):337--346.

\bibitem[Lee, 2006]{lee2006propensity}
Lee, S. (2006).
\newblock Propensity score adjustment as a weighting scheme for volunteer panel web surveys.
\newblock {\em Journal of official statistics}, 22(2):329.

\bibitem[Leemann and Wasserfallen, 2017]{leemann2017extending}
Leemann, L. and Wasserfallen, F. (2017).
\newblock Extending the use and prediction precision of subnational public opinion estimation.
\newblock {\em American journal of political science}, 61(4):1003--1022.

\bibitem[Li and Si, 2022]{li2022embedded}
Li, K. and Si, Y. (2022).
\newblock Embedded multilevel regression and poststratification: Model-based inference with incomplete auxiliary information.
\newblock {\em arXiv preprint arXiv:2205.02775}.

\bibitem[Li and Si, 2024]{li2024embedded}
Li, K. and Si, Y. (2024).
\newblock Embedded multilevel regression and poststratification: Model-based inference with incomplete auxiliary information.
\newblock {\em Statistics in Medicine}, 43(2):256--278.

\bibitem[Little et~al., 2020]{little2020measures}
Little, R.~J., West, B.~T., Boonstra, P.~S., and Hu, J. (2020).
\newblock Measures of the degree of departure from ignorable sample selection.
\newblock {\em Journal of survey statistics and methodology}, 8(5):932--964.

\bibitem[Lopez-Martin et~al., 2022]{lopez2022multilevel}
Lopez-Martin, J., Phillips, J.~H., and Gelman, A. (2022).
\newblock Multilevel regression and poststratification case studies.
\newblock {\em URL https://juanlopezmartin. github. io}, 902:903.

\bibitem[Matei, 2018]{matei2018some}
Matei, A. (2018).
\newblock On some reweighting schemes for nonignorable unit nonresponse.
\newblock {\em The Survey Statistician}, (77):21--33.

\bibitem[Matz et~al., 2017]{matz2017psychological}
Matz, S.~C., Kosinski, M., Nave, G., and Stillwell, D.~J. (2017).
\newblock Psychological targeting as an effective approach to digital mass persuasion.
\newblock {\em Proceedings of the national academy of sciences}, 114(48):12714--12719.

\bibitem[McPhee et~al., 2023]{mcpheedata}
McPhee, C., Barlas, F., Brigham, N., Darling, J., Dutwin, D., Jackson, C., et~al. (2023).
\newblock Data quality metrics for online samples: Considerations for study design and analysis.

\bibitem[Meng, 2018]{meng2018statistical}
Meng, X.-L. (2018).
\newblock Statistical paradises and paradoxes in big data (i) law of large populations, big data paradox, and the 2016 us presidential election.
\newblock {\em The Annals of Applied Statistics}, 12(2):685--726.

\bibitem[Meng, 2022]{meng2022comments}
Meng, X.-L. (2022).
\newblock Comments on" statistical inference with non-probability survey samples"-miniaturizing data defect correlation: A versatile strategy for handling non-probability samples.

\bibitem[Mercer et~al., 2018]{mercer2018weighting}
Mercer, A., Lau, A., and Kennedy, C. (2018).
\newblock For weighting online opt-in samples, what matters most?

\bibitem[Pasek and Pasek, 2018]{pasek2018package}
Pasek, J. and Pasek, M.~J. (2018).
\newblock Package ‘anesrake’.
\newblock {\em Compr. R Arch. Netw}.

\bibitem[Prosser and Mellon, 2018]{prosser2018twilight}
Prosser, C. and Mellon, J. (2018).
\newblock The twilight of the polls? a review of trends in polling accuracy and the causes of polling misses.
\newblock {\em Government and Opposition}, 53(4):757--790.

\bibitem[Rafei et~al., 2022]{rafei2022robust}
Rafei, A., Elliott, M.~R., and Flannagan, C.~A. (2022).
\newblock Robust and efficient bayesian inference for non-probability samples.
\newblock {\em arXiv preprint arXiv:2203.14355}.

\bibitem[Rivers, 2007]{rivers2007sampling}
Rivers, D. (2007).
\newblock Sampling for web surveys.
\newblock In {\em Joint Statistical Meetings}, volume~4. American Statistical Association Alexandria, VA.

\bibitem[Rocheva et~al., 2022]{rocheva2022targeting}
Rocheva, A., Varshaver, E., and Ivanova, N. (2022).
\newblock Targeting on social networking sites as sampling strategy for online migrant surveys: The challenge of biases and search for possible solutions.
\newblock {\em Migration Research in a Digitized World}, page~35.

\bibitem[Schirripa~Spagnolo et~al., 2025]{schirripa2025inference}
Schirripa~Spagnolo, F., Bertarelli, G., Summa, D., Scannapieco, M., Pratesi, M., Marchetti, S., and Salvati, N. (2025).
\newblock Inference for big data assisted by small area methods: an application on sustainable development goals sensitivity of enterprises in italy.
\newblock {\em Journal of the Royal Statistical Society Series A: Statistics in Society}, 188(1):27--45.

\bibitem[Schneider and Harknett, 2022]{schneider2022s}
Schneider, D. and Harknett, K. (2022).
\newblock What’s to like? facebook as a tool for survey data collection.
\newblock {\em Sociological Methods \& Research}, 51(1):108--140.

\bibitem[Schonlau and Couper, 2017]{10.1214/16-STS597}
Schonlau, M. and Couper, M.~P. (2017).
\newblock {Options for Conducting Web Surveys}.
\newblock {\em Statistical Science}, 32(2):279 -- 292.

\bibitem[Selcuki, 2023]{selcuki2023turkishpolls}
Selcuki, C. (2023).
\newblock Why turkish pollsters didn't foresee erdogan's win.

\bibitem[Shirani-Mehr et~al., 2018]{shirani2018disentangling}
Shirani-Mehr, H., Rothschild, D., Goel, S., and Gelman, A. (2018).
\newblock Disentangling bias and variance in election polls.
\newblock {\em Journal of the American Statistical Association}, 113(522):607--614.

\bibitem[Si, 2020]{si2020use}
Si, Y. (2020).
\newblock On the use of auxiliary variables in multilevel regression and poststratification.
\newblock {\em arXiv preprint arXiv:2011.00360}.

\bibitem[Smith, 1976]{smith1976foundations}
Smith, T. (1976).
\newblock The foundations of survey sampling: a review.
\newblock {\em Journal of the Royal Statistical Society: Series A (General)}, 139(2):183--195.

\bibitem[Sohlberg et~al., 2017]{sohlberg2017determinants}
Sohlberg, J., Gilljam, M., and Martinsson, J. (2017).
\newblock Determinants of polling accuracy: the effect of opt-in internet surveys.
\newblock {\em Journal of Elections, Public Opinion and Parties}, 27(4):433--447.

\bibitem[Stephan, 1942]{stephan1942iterative}
Stephan, F.~F. (1942).
\newblock An iterative method of adjusting sample frequency tables when expected marginal totals are known.
\newblock {\em The Annals of Mathematical Statistics}, 13(2):166--178.

\bibitem[Sturgis et~al., 2018]{sturgis2018assessment}
Sturgis, P., Kuha, J., Baker, N., Callegaro, M., Fisher, S., Green, J., Jennings, W., Lauderdale, B.~E., and Smith, P. (2018).
\newblock An assessment of the causes of the errors in the 2015 uk gseneral election opinion polls.
\newblock {\em Journal of the Royal Statistical Society Series A: Statistics in Society}, 181(3):757--781.

\bibitem[Tam and Clarke, 2015]{tam2015big}
Tam, S.-M. and Clarke, F. (2015).
\newblock Big data, official statistics and some initiatives by the australian bureau of statistics.
\newblock {\em International Statistical Review}, 83(3):436--448.

\bibitem[Tan, 2007]{tan2007comment}
Tan, Z. (2007).
\newblock Comment: Understanding or, ps and dr.
\newblock {\em Statistical Science}, 22(4):560--568.

\bibitem[Tutz, 2023]{tutz2023probability}
Tutz, G. (2023).
\newblock Probability and non-probability samples: Improving regression modeling by using data from different sources.
\newblock {\em Information Sciences}, 621:424--436.

\bibitem[Unangst et~al., 2020]{unangst2020process}
Unangst, J., Amaya, A.~E., Sanders, H.~L., Howard, J., Ferrell, A., Karon, S., and Dever, J.~A. (2020).
\newblock A process for decomposing total survey error in probability and nonprobability surveys: A case study comparing health statistics in us internet panels.
\newblock {\em Journal of Survey Statistics and Methodology}, 8(1):62--88.

\bibitem[Valentino et~al., 2020]{valentino2020consequences}
Valentino, N.~A., Zhirkov, K., Hillygus, D.~S., and Guay, B. (2020).
\newblock The consequences of personality biases in online panels for measuring public opinion.
\newblock {\em Public Opinion Quarterly}, 84(2):446--468.

\bibitem[Valliant, 2019]{Vaillant_2019}
Valliant, R. (2019).
\newblock {Comparing Alternatives for Estimation from Nonprobability Samples}.
\newblock {\em Journal of Survey Statistics and Methodology}, 8(2):231--263.

\bibitem[Valliant, 2020]{valliant2020comparing}
Valliant, R. (2020).
\newblock Comparing alternatives for estimation from nonprobability samples.
\newblock {\em Journal of Survey Statistics and Methodology}, 8(2):231--263.

\bibitem[Valliant et~al., 2013]{valliant2013practical}
Valliant, R., Dever, J.~A., and Kreuter, F. (2013).
\newblock {\em Practical tools for designing and weighting survey samples}, volume~1.
\newblock Springer.

\bibitem[Wang et~al., 2015]{wang2015forecasting}
Wang, W., Rothschild, D., Goel, S., and Gelman, A. (2015).
\newblock Forecasting elections with non-representative polls.
\newblock {\em International Journal of Forecasting}, 31(3):980--991.

\bibitem[West and Andridge, 2023]{west2023evaluating}
West, B.~T. and Andridge, R.~R. (2023).
\newblock Evaluating pre-election polling estimates using a new measure of non-ignorable selection bias.
\newblock {\em Public Opinion Quarterly}, 87(S1):575--601.

\bibitem[Wiegand, 1968]{wiegand1968kish}
Wiegand, H. (1968).
\newblock Kish, l.: Survey sampling. john wiley \& sons, inc., new york, london 1965, ix+ 643 s., 31 abb., 56 tab., preis 83 s.

\bibitem[Wu, 2022]{wu2022statistical}
Wu, C. (2022).
\newblock Statistical inference with non-probability survey samples.
\newblock {\em Surv. Methodol}, 48:283--311.

\bibitem[Zagheni et~al., 2014]{zagheni2014inferring}
Zagheni, E., Garimella, V. R.~K., Weber, I., and State, B. (2014).
\newblock Inferring international and internal migration patterns from twitter data.
\newblock In {\em Proceedings of the 23rd international conference on world wide web}, pages 439--444.

\bibitem[Zagheni and Weber, 2015]{zagheni2015demographic}
Zagheni, E. and Weber, I. (2015).
\newblock Demographic research with non-representative internet data.
\newblock {\em International Journal of Manpower}, 36(1):13--25.

\bibitem[Zagheni et~al., 2017]{zagheni2017leveraging}
Zagheni, E., Weber, I., and Gummadi, K. (2017).
\newblock Leveraging facebook's advertising platform to monitor stocks of migrants.
\newblock {\em Population and Development Review}, pages 721--734.

\bibitem[Zarouali et~al., 2022]{zarouali2022using}
Zarouali, B., Dobber, T., De~Pauw, G., and de~Vreese, C. (2022).
\newblock Using a personality-profiling algorithm to investigate political microtargeting: assessing the persuasion effects of personality-tailored ads on social media.
\newblock {\em Communication Research}, 49(8):1066--1091.

\bibitem[Zou et~al., 2016]{zou2016variance}
Zou, B., Zou, F., Shuster, J.~J., Tighe, P.~J., Koch, G.~G., and Zhou, H. (2016).
\newblock On variance estimate for covariate adjustment by propensity score analysis.
\newblock {\em Statistics in medicine}, 35(20):3537--3548.

\bibitem[Łukasz Chrostowski and Beręsewicz, 2024]{nonprobsvy}
Łukasz Chrostowski and Beręsewicz, M. (2024).
\newblock {\em nonprobsvy: Inference Based on Non-Probability Samples}.
\newblock R package version 0.1.0.

\end{thebibliography}

\end{document}